\documentclass[12pt,a4paper]{article}

\usepackage{amssymb,amsmath,amsfonts,eurosym,geometry,graphicx,color,comment,footmisc,caption,pdflscape,subcaption,array,makecell,ragged2e, arydshln, titlesec, float, hyperref}
\usepackage{ulem}
\usepackage{sectsty}

\usepackage[style=apa, backend=biber]{biblatex}
\DeclareDelimFormat{finalnamedelim:parencite}{\addspace\&\space}
\DeclareDelimFormat{finalnamedelim:textcite}{\addspace and\space}
\DeclareDelimFormat{multinamedelim}{\addcomma\space}          
\DeclareDelimFormat{multinamedelim:textcite}{\addcomma\space} 

\usepackage{caption}
\usepackage{setspace}   

\DeclareCaptionLabelFormat{plain}{#1~#2} 
\newcolumntype{L}[1]{>{\raggedright\let\newline\\arraybackslash\hspace{0pt}}m{#1}}
\newcolumntype{C}[1]{>{\centering\let\newline\\arraybackslash\hspace{0pt}}m{#1}}
\newcolumntype{R}[1]{>{\raggedleft\let\newline\\arraybackslash\hspace{0pt}}m{#1}}

\hypersetup{hidelinks}

\begin{document}

\begin{titlepage}
\title{Trimming of extreme votes and favoritism: \\ Evidence from the field\thanks{This research was funded by the German Research Foundation (DFG project number 525061914). Ethical approval for this study was received by the Institutional Review Board (IRB) at the Faculty of Economics and Social Sciences of the University of Tübingen under the registry number A2.5.4-322\_hb. Generative AI was employed solely for the purpose of refining the linguistic quality of the manuscript. All conceptual development, data analysis, and final interpretations remain the original work of the authors.}}
\author{\normalsize Alex Krumer\thanks{Molde University College, alex.krumer@himolde.no} \and \normalsize Felix Otto\thanks{Corresponding author. University of Tübingen, felix.otto@uni-tuebingen.de} \and \normalsize Tim Pawlowski\thanks{University of Tübingen, tim.pawlowski@uni-tuebingen.de}}
\date{\small \today}
\maketitle
\begin{abstract}
\noindent
\begin{spacing}{1}
\begin{quote}

\footnotesize Despite a large body of theoretical literature on voting mechanisms, there is no documented evidence from real-world panel evaluations about the effect of trimming the extreme votes on sincere voting. We provide the first such evidence by comparing subjective evaluations of experts from different countries in competitive settings with and without a trimming mechanism. In these evaluations, some of the evaluated subjects are experts’ compatriots. Using data on 29,383 subjective evaluations, we find that experts assign significantly higher scores to their compatriots in panels without trimming. However, in panels with trimming, this favoritism is generally insignificant.\\

\vspace{0.2in}
\noindent\justifying\textbf{Keywords:} Subjective performance evaluation, Judging panel, Jury, Favoritism, Nationalistic bias, Trimming

\vspace{0in}
\end{quote}
\bigskip
\end{spacing}
\end{abstract}
\setcounter{page}{0}
\thispagestyle{empty}
\end{titlepage}
\pagebreak \newpage

\onehalfspacing

\section{Introduction} \label{sec:introduction}
\noindent Subjective performance evaluations are commonly used by organizations to assess, select, and rank candidates in different competitive situations; for instance, awarding research grants, selecting candidates in the labor market, choosing the best wine in a wine competition, or judging the performance of athletes in sports competitions. Since many such evaluations have serious economic consequences for both candidates and organizations, the common practice to ensure the quality of evaluation is to appoint panels of experts.

One seemingly innocuous assumption in such evaluations is that the experts make their evaluations sincerely. In fact, sincere voting is backed up by two seminal theoretical papers. \textcite{black1948} established a median voter theorem according to which the best strategy is approaching a median voter preference if voter’s preferences are single-peaked. \textcite{moulin1980} showed that in the case of such single-peaked preferences, sincere voting is a dominant strategy under the median rule (i.e., the selected alternative is the one that was selected by the median voter).

However, most real-world preferences are rather multi-dimensional than single-peaked, and the decision-making mechanism rarely follows the median rule. Thus, despite our desire to believe in a just and fair world, the assumption of sincere voting is unfortunately too strong, especially in settings where individuals may have conflicting interests. This claim is backed up by \textcite{austensmith1996} who showed that sincere voting is not rational, even when individuals have a common preference to choose the best alternative. One example of deviation from sincere evaluation is favoritism, that is driven by certain similarities between the experts and the subjects of their evaluations.

In fact, there is substantial evidence from a variety of settings that has documented favoritism by judges on panels. For instance, it has been shown that applicants for research grants have a higher probability of success if they have the same affiliation as one of the judges of the evaluation panel \parencite[e.g.,][]{kwon2021, sandstrom2008}. In line with this, there is a voting bias in the Eurovision Song Contest that is driven by geographical or cultural proximities \parencite[e.g.,][]{ginsburgh2008, spierdijk2009}. In international sports competitions, judges tend to assign higher scores to their compatriots in ski jumping, gymnastics, diving, figure skating, and dressage \parencite{emerson2009, heiniger2021, krumer2022, sandberg2018, zitzewitz2006}.\footnote{Favoritism may also be driven by other characteristics such as similarity in race \parencite{price2010}, ethnicity \parencite{shayo2011}, or language \parencite{faltings2023}.} These findings raise concerns about the quality of evaluations and are worrisome for candidates and organizations alike.

Given the widespread and well-documented phenomenon of favoritism, it is natural to look for effective measures to mitigate it in panel evaluations. A frequently applied remedy for such biases is trimming (often described as Olympic mean). The simple and straightforward approach of trimming is to exclude the most extreme scores from the final panel evaluation---typically the highest (or several highest) and lowest (or several lowest) scores---using the mean score among the remaining ones. In its extreme form, i.e., when trimming reaches 50\% from both ends of the data, the trimmed mean reflects the median rule.

Trimming mechanisms have been widely used, for instance, in determining interest rates such as the London Interbank Offered Rate (LIBOR) that used to be the key global benchmark for borrowing between banks \parencite{eisl2017}, or for performance evaluations in different sports such as figure skating and ski jumping \parencite{zitzewitz2006}. The main reason for the usage of trimming in panel evaluations seems intuitively clear: by removing scores that potentially contain the largest degree of bias, trimming should mitigate any bias \emph{ex post}, i.e., after the evaluators assign their scores. But does it also contribute to more sincere voting \emph{ex ante}? In other words, do individual evaluators behave differently with and without trimming?

The International Organization of Securities Commissions (IOSCO) recommends using trimmed means because this mechanism ``\emph{reduce[s] the individual incentive to try to manipulate}'' \parencite[p. 18]{iosco2013}. A plausible reason supporting this claim could be reputational concerns: for an evaluator to be perceived as unbiased, the assigned score should be in the subset of non-trimmed scores. This is more likely to happen if the assigned score reflects the actual performance of the candidate. However, an evaluator that wishes to favor her or his preferred candidate should assign the highest score (which will obviously be trimmed but still increase the overall average by keeping the second highest score of another evaluator within the subset of the remaining scores). This illustrates a trade-off between private preferences (i.e., individual’s internal desires) and reputational preferences (i.e., gaining social approval or avoiding negative judgment) of the evaluators that act in different directions \parencite{miceli1994}. Moreover, even under single-peaked preferences, there can be several Nash equilibria, one of which includes non-sincere voting \parencite{puppe2021}. Therefore, it is theoretically unclear whether trimming the extreme votes reduces bias of individual evaluators.

To develop evidence-based policies for combating biases in (so widely used) evaluation panels, empirical evidence on the effects of trimming on individual voting is needed. There are two notable papers using lab experiments to observe actual voting patterns with and without a trimming mechanism. \textcite{puppe2021} find a significantly higher share of sincere voting under extreme trimming (i.e., median rule) compared to no trimming. \textcite{louis2023} find that individuals always misreport, but, conditional on being dishonest, they report values closer to their optimal ones as the degree of trimming increases.
\clearpage
While these experiments confirmed the theoretical prediction that under extreme trimming, voters behave closer to sincerity,\footnote{We also refer the reader to theoretical contributions of \textcite{gruner2004} and \textcite{rosar2015} who compared between extreme trimming (i.e., median) and average mechanisms.} the experimental designs intended to resemble the allocation of budget between different projects. This is because each participant in both papers had different valuations (peaks) regarding several public projects. However, such a design differs from the widely used evaluation processes of individual candidates, where evaluators are supposed to have similar and unbiased voting criteria when assessing the same candidates. Additionally, while laboratory experiments provide valuable insights, their methodology is not beyond critique. For example, according to Robert Aumann ``\emph{in experiments … the monetary payoff is usually very small. More importantly, the decisions that people face are not ones that they usually take, with which they are familiar. The whole setup is artificial. It is not a decision that really affects them and to which they are used}'' \parencite[p. 712]{hart2005}.

Thus, we need evidence from real-world settings. Surprisingly, however, despite the widespread evidence of favoritism in subjective evaluations, and the very well-developed theoretical literature on a variety of voting mechanisms,\footnote{See for example, \textcite{balinski2007}, \textcite{dasgupta2020}, \textcite{gershkov2017}, \textcite{kerman2024}, \textcite{kleiner2017}, and references therein.} there is not a single study using real-world data to investigate the effect of trimming on favoritism (or any other bias) of individual evaluators.\footnote{It is worth noting the study by \textcite{eisl2017} who created a hypothetical scenario as if no trimming was used in determining the LIBOR rate. It shows that untrimmed mean would lead to more manipulation compared to the trimmed one. However, that paper never observed cases without trimming. Thus, the behavior of banks could differ if they actually operate in a setting without trimming.} The perfect way to measure the effect of trimming on favoritism would be to observe evaluations of individuals who are prone to bias on panels \emph{without} a trimming mechanism and compare it to evaluations of individuals who are prone to bias on panels \emph{with} trimming. The likely reason for the absence of such real-world evidence is that it is almost impossible to find a natural setting with an a-priori well-defined group of evaluators who are plausibly suspected of being biased and whose evaluations are observed for similar tasks with and without a trimming mechanism.

In this paper, we overcome these obstacles by using a unique setting where we are able to compare the evaluations of professional experts who assess the performance of highly incentivized and professional individuals in real competitive settings with and without trimming the extreme evaluation scores. More precisely, we use the Olympic sport of freeski halfpipe where professional athletes perform a series of jumps and tricks on their skis while going down the length of a large halfpipe made of snow.\footnote{See \textcite{bareli2020} and a recent comprehensive review by \textcite{palacioshuerta2025} for advantages to use sports data for economic research.} After they complete their performance, each member (judge) of a jury panel assigns a score.\footnote{See Section 2, for additional details on freeski halfpipe.} One of these judges can be a compatriot of the performing athlete, making this judge a perfect subject to study (potentially) biased evaluations.

Most importantly, up to the 2018/19 season the scores of all judges on the jury panel were used to calculate the average without trimming the extreme scores. However, starting from the 2018/19 season, the highest and lowest scores were trimmed. This change allows us to compare individual evaluations of judges with and without the trimming mechanism in order to test whether judges assign higher scores to their compatriots and whether such favoritism varies between both periods. Despite this setting being unusual, it has the advantage of providing clean identification and is therefore a ``\emph{perfect domain}'' \parencite[p. 45]{list2020} for our research question.

We use data on 29,383 subjective evaluations from the most prestigious competitions. Controlling for within-performance and within-judge variation, our difference-in-differences analyses show a significant nationalistic bias before the change according to which judges assign significantly higher scores to their compatriots. We also find that, before the introduction of the trimming mechanism, judges assigned the highest score to their compatriots significantly more often than the unbiased share in the absence of any bias (which should be 20\% in a panel of five judges, for instance). However, after the introduction of the trimming mechanism, the nationalistic bias became much lower and, in most cases, insignificant, suggesting that the trimming mechanism has a positive effect on the sincerity of subjective evaluations of judges who are prone to nationalistic bias. This finding supports the effectiveness of the rather simple method of trimming, which even seems to improve evaluations of highly qualified experts.

\section{Setting} \label{sec:setting}
Freeski halfpipe competitions are staged under the umbrella of the Fédération Internationale de Ski (FIS), which is the international governing body for skiing and snowboarding.\footnote{Do not confuse with snowboard halfpipe that has always used the trimming method.} Starting from the 2014 Winter Olympics Games, freeski halfpipe is an official Olympic medal event.

In freeski halfpipe, athletes perform a series of jumps and tricks on their skis while going down the length of a large halfpipe made of snow.\footnote{To better understand the freeski halfpipe sport, the interested reader can watch some examples of the top performances on \url{https://www.youtube.com/watch?v=GYDCeih-cac}. Last accessed on 11/12/2025.} This is called a ‘run’ and defines an athlete’s performance. Men’s and women’s competitions consist of a qualification phase in which athletes usually perform two runs and a final phase with two or three runs. In both phases, only the best run counts for the final ranking. Depending on the size of the competitor field, only between 8 and 12 top ranked athletes from the qualification phase qualify for the final.
\clearpage
The quality of freeski performances is determined entirely by the evaluations of a jury panel. Each judge assigns a single overall score in the range of 0 to 100 (integer) for each run. The judges’ scores are then averaged and shown to two decimal places to determine the total run score. Judges evaluate each run independently based on guidelines that include five criteria (execution, difficulty, amplitude, variety, and progression), while considering falls, stops, or mistakes. It is formally specified that judges must be fair and unbiased, explicitly regarding national affiliation, race, gender, or sponsor. However, judges are also entitled to add subjectivity and personal preference to their evaluation, for example, to better compare between two runs or the execution of the same trick \parencite{fis2019a}. To prevent extreme discrepancies among judges’ evaluations, a head judge is overseeing the judges’ decisions and scoring results. This process was manually managed up to the 2019/20 season. Since then, it has been assisted by a scoring gap alert that is activated in case of extreme deviation between adjacent judges’ scores. While there are no direct sanctions, it is plausible to assume that judges with frequently extreme opinions could face future career consequences.\footnote{While extreme deviations among judges’ scores are rather unusual, we nonetheless observe in our data that judges use a considerable range of scores to evaluate performances throughout all seasons. For instance, while the average distance of scores from each other for a given performance is 2.40 points (SD=1.56), the maximum distance of scores assigned for a run is 14.47 points that was given at a competition in 2022. It is also important to note that the within-performance standard deviation did not become smaller with the automatic gap alert.}

Judges are officially qualified and licensed experts and appointed by the FIS Council on recommendation of the Snowboard Freestyle Freeski Committee prior to the upcoming seasons. This process should ensure that only well-trained and competent judges are selected for this task. In addition, according to FIS regulations, all judges on a jury panel must represent a different country \parencite{fis2019b}. While most judges are men, many competitions also have at least one female judge.

\begin{figure}[!htp]
    \centering
    \caption{Example of competition results without and with trimming as presented in the official results documents}
    \label{fig1}
    \includegraphics[width=1\linewidth]{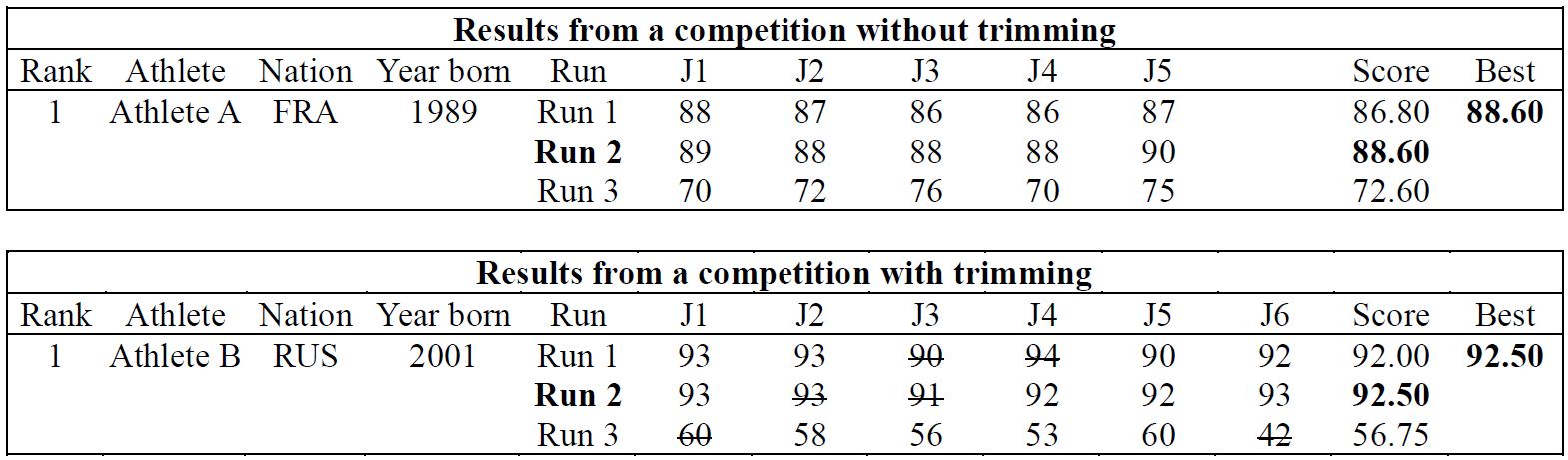}
    \caption*{\footnotesize\justifying\textit{Notes:} $J_i$ represents different judges.}
\end{figure}

The most important feature that directly relates to our study is that up to the 2018/19 season there were five judges on the jury panel whose scores were used to calculate the average without trimming the extreme scores. However, starting from the 2018/19 season, the FIS switched to six (and, in a few cases, to seven) judges on the jury panel trimming the highest and lowest scores. In both periods, before and after the change, the judges’ identities and their scores are fully observable. After the introduction of the trimming method, the highest and lowest scores are marked as deleted (see Figure~\ref{fig1}), so everyone can see instantly which scores were trimmed. In case the highest (lowest) score was awarded by more than one judge (e.g., Run 2 of Athlete B in Figure~\ref{fig1}), the deleted score is determined randomly.

\section{Data and descriptive statistics} \label{sec:data}
\subsection{Data collection and sample}
We collected data from the official website of the FIS on all men’s and women’s World Cups, World Championships, and Olympic Games for the seasons between 2012/13 and 2023/24.\footnote{As for most international sports, the COVID-19 pandemic also affected the freeski calendar. While competitions took place in all years, there were only a few events in the 2020/2021 season, which we do not include in the dataset due to irregularities in the competition format.} These are the most prestigious competitions in professional freeski halfpipe. The data includes information on judge and athlete characteristics (i.e., name-IDs, gender, and country), as well as event and performance characteristics (i.e., event type, date, location, and competition phase). We excluded competitions with inconsistencies in the competition or in the judging panel composition\footnote{These are events for which only qualification results were available, events with missing or incomplete information on performance evaluations, and events where more than one judge in the panel comes from the same country or where the organizers were not able to employ six judges after the rule change (e.g., if judges got sick or due to travel restrictions during the COVID-19 pandemic).} as well as qualifications where split panels and no trimming was used.\footnote{This so called double-up heat format was only used for qualifications with a large competitor field.}

\begin{table}[!htbp]
\centering
\footnotesize
\caption{Sample size}
\label{tab1}
\begin{tabular}{lccc} 
\hline\hline
 & \makecell{Without trimming \\ (2013-2018 seasons)}
 & \makecell{With trimming \\ (2019-2024 seasons)}
 & \makecell{Total sample \\ (all seasons)}
 \\ 
\hline
No. of men’s competitions & 18 & 18 (2) & 36 \\
No. of women’s competitions & 19 & 18 (2) & 37 \\
No. of male athletes & 109 & 91 & 165 \\
No. of female athletes & 77 & 53 & 106 \\
No. of athlete countries & 31 & 20 & 33 \\
No. of judges & 29 & 38 & 55 \\
No of judge countries & 13 & 15 & 16 \\
No. of runs & 2,739 & 2,566 (292) & 5,305 \\
No of scores & 13,695 & 15,688 (2,044) & 29,383 \\
\hline\hline
\end{tabular}
\vspace{1mm}
\parbox{\linewidth}{\vspace{1mm}\footnotesize\noindent\justifying\textit{Notes:} Number of observations with seven (instead of six) judges in the panel with trimming are shown in parentheses.}
\end{table}

As shown in Table~\ref{tab1}, our final dataset consists of 36 men’s and 37 women’s competitions. It includes more than 250 different athletes from more than 30 different countries that were evaluated by 55 different judges from 16 different countries. Overall, we have 2,739 performances from men’s and women’s competitions before the rule change and 2,566 performances after the trimming mechanism was introduced. This translates into 29,383 individual judges’ scoring decisions on 5,305 performances.

\subsection{Descriptive statistics}

We structure the panel dataset at the level of judges’ scores to compare evaluations between and within athlete performances and individual judges. This structure allows us to follow judges over time and compare evaluations of judges in different decision-making situations, i.e., when athletes are compatriots or non-compatriots as well as with and without trimming.

Table~\ref{tab2} provides summary statistics of judges’ scores for compatriot and non-compatriot judges as well as before and after the introduction of trimming. We see that before the change; judges assign on average 0.43 more points in men’s and 0.58 more points in women’s competitions when evaluating compatriot athletes compared to the other judges on the panel. However, this gap has narrowed after the change, where compatriot judges assign on average 0.10 fewer points than the other judges on the panel in men’s competitions and only 0.21 more points in women’s competitions. Moreover, athletes without a compatriot judge on the panel receive on average lower scores than their competitors with a compatriot on the panel in both periods.

\begin{table}[!htb]
\centering
\footnotesize
\caption{Descriptive statistics: trimming and being a compatriot judge}
\label{tab2}
\resizebox{\textwidth}{!}{
\begin{tabular}{lcccccc}
\hline\hline
 & \multicolumn{3}{c}{\makecell{\rule{0pt}{4mm}Without trimming \\ (2013-2018 seasons)}} 
 & \multicolumn{3}{c}{\makecell{\rule{0pt}{4mm}With trimming \\ (2019-2024 seasons)}} \\
\hline
 & \makecell{\rule{0pt}{4mm}Compatriot \\ judges} 
 & \makecell{\rule{0pt}{4mm}Non-compatriot \\ judges on \\ same panels} 
 & \multicolumn{1}{c|}{\makecell{\rule{0pt}{4mm}Judges on \\ panels without \\ compatriot}} 
 & \makecell{\rule{0pt}{4mm}Compatriot \\ judges} 
 & \makecell{\rule{0pt}{4mm}Non-compatriot \\ judges on \\ same panels} 
 & \makecell{\rule{0pt}{4mm}Judges on \\ panels without \\ compatriot} \\
\hline\hline
 & \multicolumn{6}{c}{\rule{0pt}{4mm}Men's competitions} \\
\hline
Judge scores        &            &              & \multicolumn{1}{c|}{}              &            &              &              \\
Mean                & 56.94      & 56.51        & \multicolumn{1}{c|}{49.04}        & 53.83      & 53.93        & 51.24        \\
Overall \textbar{} within SD & 28.10 \textbar{} NA & 28.26 \textbar{} 2.01 & \multicolumn{1}{c|}{25.03 \textbar{} 2.24} & 28.34 \textbar{} NA & 28.11 \textbar{} 2.07 & 28.60 \textbar{} 2.32 \\
Min-max             & 1-98       & 1-98         &\multicolumn{1}{c|}{1-98}         & 2-98      & 1-98         & 1-98         \\
\hline
\makecell[l]{No. of score obs. \\ (runs)} &
\makecell{797 \\ (797)} &
\makecell{3,188 \\ (797)} &
\multicolumn{1}{c|}{\makecell{3,965 \\ (793)}} &
\makecell{868 \\ (868)} &
\makecell{4,419 \\ (868)} &
\makecell{3,943 \\ (642)} \\
\hline\hline
 & \multicolumn{6}{c}{\rule{0pt}{4mm}Women's competitions} \\
\hline
Judge scores        &            &              &\multicolumn{1}{c|}{}              &            &              &              \\
Mean                & 61.39      & 60.81        &\multicolumn{1}{c|}{54.51}        & 60.17      & 59.96        & 59.41        \\
Overall \textbar{} within SD & 25.08 \textbar{} NA & 25.10 \textbar{} 2.07 & \multicolumn{1}{c|}{22.43 \textbar{} 2.36} & 25.59 \textbar{} NA & 25.60 \textbar{} 2.09 & 25.78 \textbar{} 2.35 \\
Min-max             & 2-93       & 1-96         &\multicolumn{1}{c|}{1-96}         & 3-96       & 1-96         & 1-99         \\
\hline
\makecell[l]{No. of score obs. \\ (runs)} &
\makecell{492 \\ (492)} &
\makecell{1,968 \\ (492)} &
\multicolumn{1}{c|}{\makecell{3,285 \\ (657)}} &
\makecell{493 \\ (493)} &
\makecell{2,506 \\ (493)} &
\makecell{3,459 \\ (563)} \\
\hline\hline
\end{tabular}
}
\end{table}

With scores between 1 and 99, we also see that judges use almost the full range of scores in their evaluations. In addition, we see that the within-performance standard deviation of runs with a compatriot judge on panel remains fairly similar before and after the introduction of the trimming mechanism for both genders (even with a slight increase for men). For performances with no compatriot on panel, this within-performance standard deviation also remains the same in the case of women. However, it slightly increased in men’s competitions with no compatriot judge on panel, showing that variation of judges’ scores for the same performances increased.

It is also worth noting that the average scores in women’s competitions are higher than for men. A plausible explanation for this difference could be that, on average, men might attempt more difficult maneuvers and take higher risks, which leads to more mistakes and lower scores. This intuition seems to be supported by our data. For example, if we compare the share of very low scores (below 25 points) between men’s and women’s competitions, we find that this share is 22\% in men's, but only 14\% in women's competitions.

\section{Empirical evidence} \label{sec:result}
\subsection{Distribution of the score ranks}
To explore the effect of trimming the extreme scores on the size of nationalistic bias, we first rank the scores of the judges within each run if they evaluated their compatriots. We follow the idea presented by \textcite{wolfers2006}, who showed graphically that the distribution of real scores in collegiate basketball games differed significantly from the theoretically unbiased distribution for the scores that are close to the betting threshold. \textcite{wolfers2006} used this figure as evidence of illegal point shaving in basketball games.

Following a similar logic, in the absence of nationalistic bias, we would expect to see a uniform distribution of the share of the rankings of the scores (from the highest to the lowest score) from a judge who evaluates her or his compatriots. This means that before the introduction of the trimming mechanism, in the panel of five judges, the unbiased share of the highest scores assigned by the judge who evaluates her or his compatriots should be 20\%. After the change, the panel of judges increased from five to six. Thus, the unbiased share of the highest scores awarded to compatriot athletes should be 16.7\%.\footnote{Note that it is possible that several judges can assign similar scores. Thus, among the judges who assigned the same score, we split their rankings randomly (into the highest, second highest, etc.). This randomization resembles the rule of trimming presented in Figure~\ref{fig1}.}

\begin{figure}[H]
  \centering
  \caption{Score level distribution for compatriot judges}
  \label{fig2}
    \begin{subfigure}{0.49\textwidth}
    \fbox{\includegraphics[width=\linewidth]{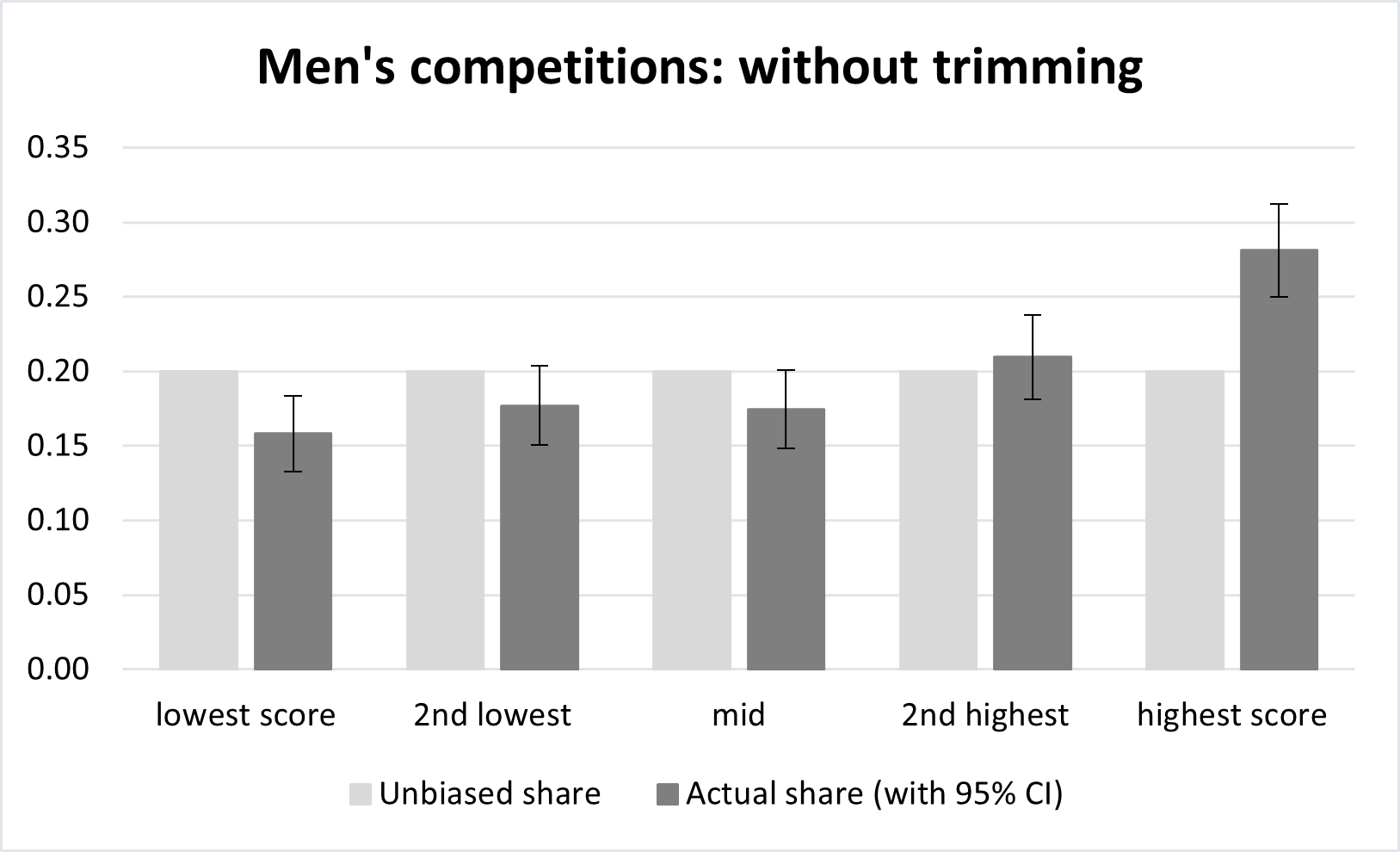}}
  \end{subfigure}
  \begin{subfigure}{0.49\textwidth}
    \fbox{\includegraphics[width=\linewidth]{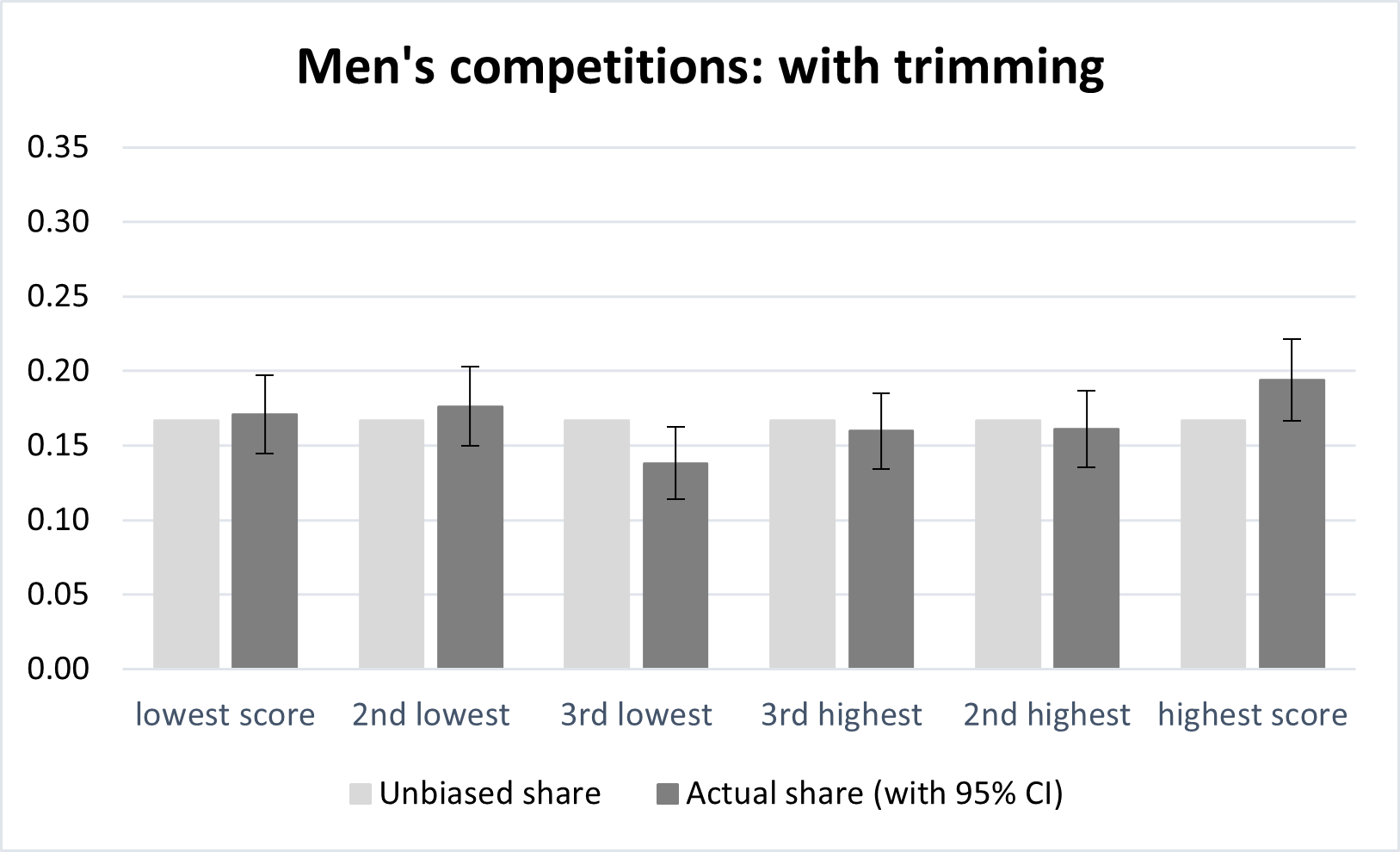}}
  \end{subfigure}
\vspace{0.2em}
  \begin{subfigure}{0.49\textwidth}
    \fbox{\includegraphics[width=\linewidth]{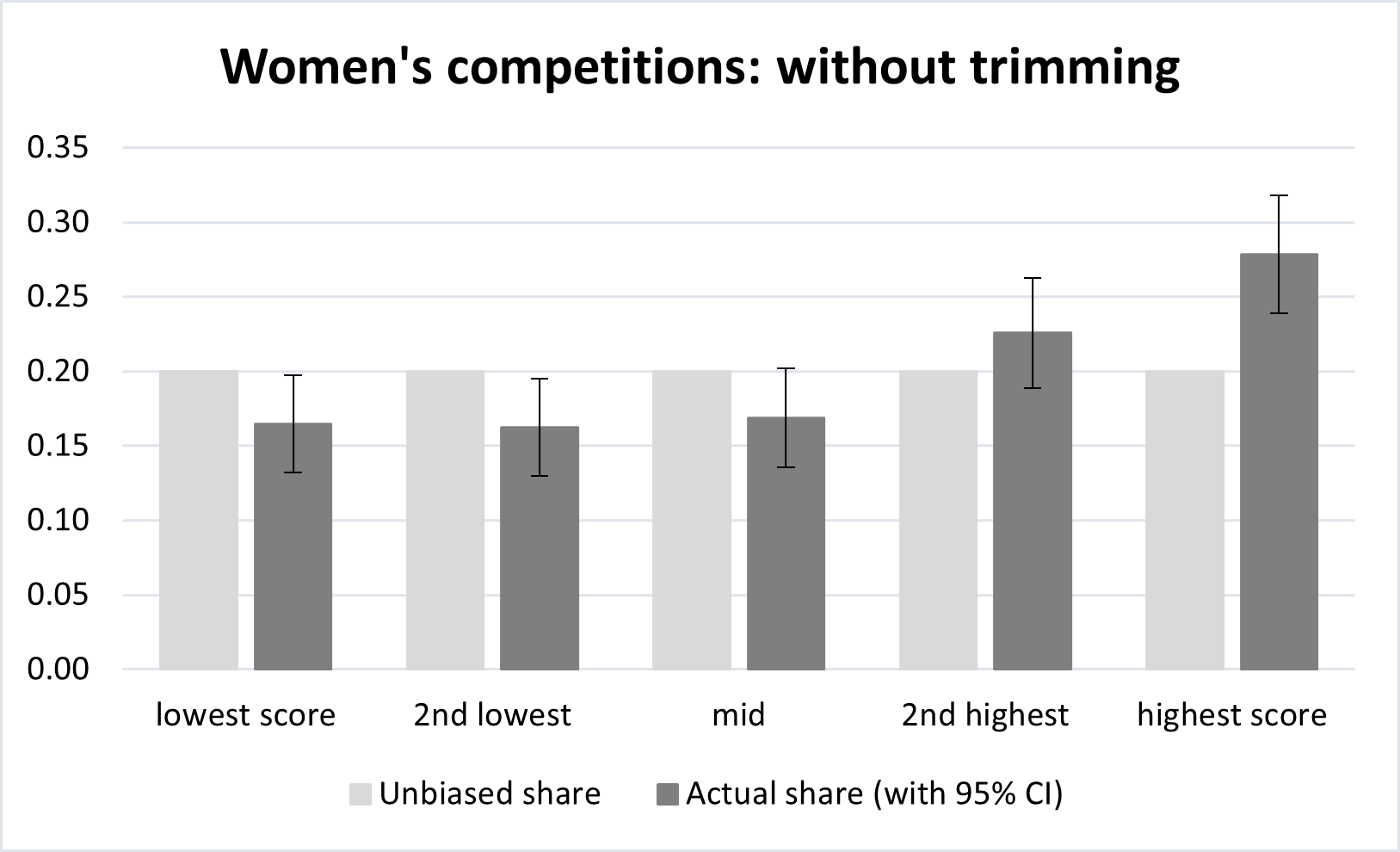}}
  \end{subfigure}
  \begin{subfigure}{0.49\textwidth}
    \fbox{\includegraphics[width=\linewidth]{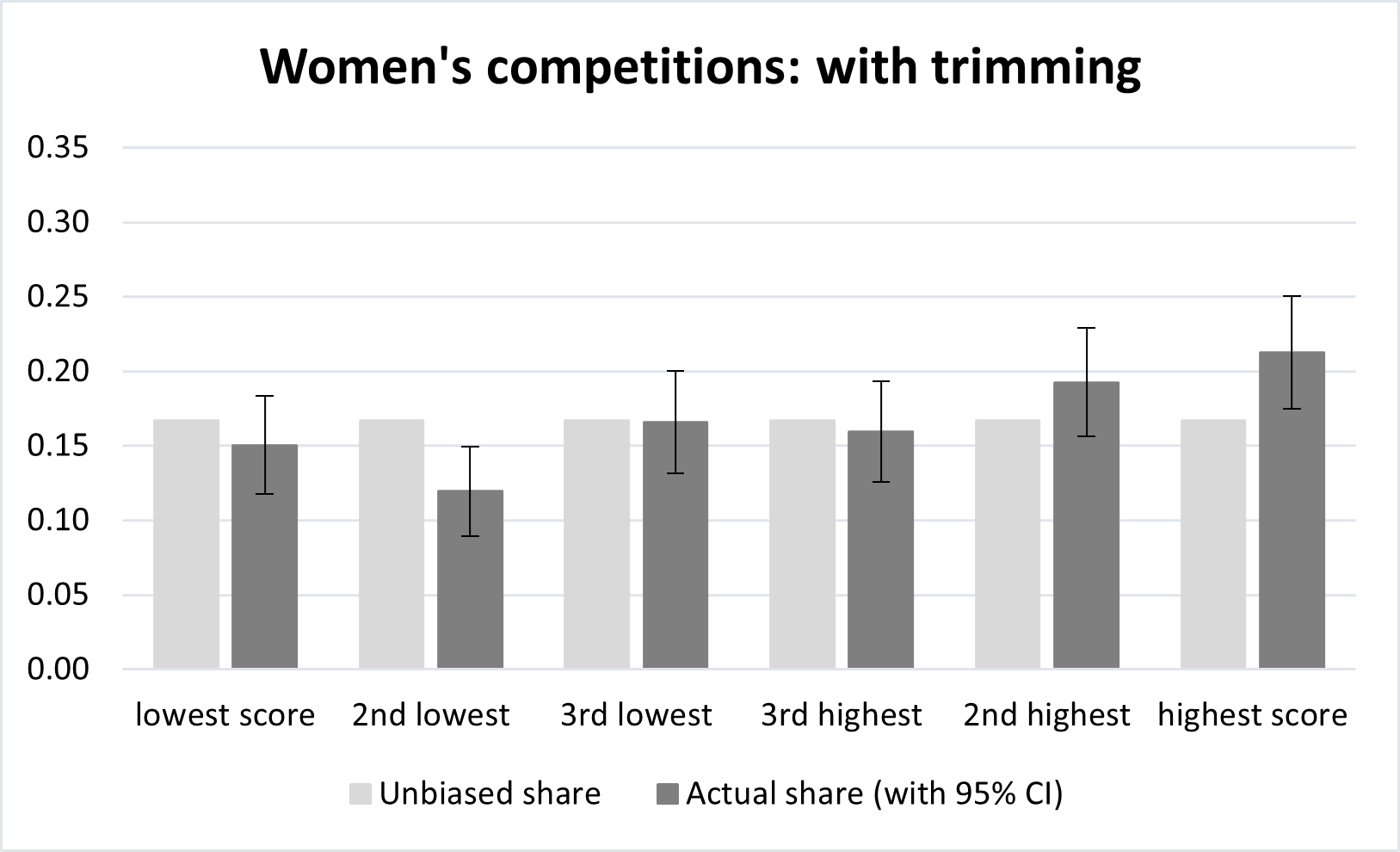}}
  \end{subfigure}
    \caption*{\footnotesize\justifying\textit{Notes:} Competitions with seven judges in the panel were removed from the sample for this comparison.}
\end{figure}

Figure~\ref{fig2}  presents, separately for both genders, the distribution of the actual scores compared to the unbiased distribution for the cases when a judge evaluated her or his compatriot without and with the trimming mechanism. We see that before the change, the shares of the highest scores assigned to compatriot performances in men’s and women’s competitions are 28.1\% and 27.9\%, respectively, which is significantly higher than the unbiased share of 20\%. More precisely, it is 41\% and 39\% above the unbiased shares for men and women, respectively. After the change, the share of the highest scores in men’s competitions is reduced to 19.4\%, which is still 16\% above the unbiased share. However, the 95\% confidence interval overlaps with this unbiased share. In women’s competitions, when judges evaluate their compatriots, they still assign the highest score among the panel members in 21.2\% of cases. This is 27\% higher than the unbiased share and statistically different from it at the 5\% level.

We also see that before the change, the share of the lowest scores assigned to compatriot athletes is significantly lower than 20\% in both men’s (15.8\%) and women’s (16.5\%) competitions. After the change, these shares do not differ significantly from the unbiased share of 16.7\% in both competitions. However, the share of the second lowest score (12\%) is still significantly lower than the unbiased share in the women’s competitions.

Overall, the results presented in Figure~\ref{fig2} serve as first evidence of the existence of nationalistic bias before and its mitigation after the introduction of the trimming mechanism. A possible reason for the remaining bias in women’s competitions may be that women emphasize more style, artistry, and overall presentation, giving rise to subjectivity in evaluations. In contrast, raw power and amplitude are more likely in men’s competitions that often involve riskier elements (e.g. tricks like spins with 1620 degrees of rotation) which leave less scope for interpretation (if successfully performed).\footnote{See an example of a spin with 1620 degrees rotation on  \url{https://www.youtube.com/watch?v=7R_n_NepAqs}. Last accessed on 11/12/2025.} While we are unable to test this claim empirically, it is in line with \textcite[p. 79]{zitzewitz2006} who noted that ``\emph{biases are larger where scoring is more subjective, as it is for ice dancing, where skaters do not have as many mandatory deductions for falls, and for artistic impression as opposed to technical merit scores}''. It is also in line with \textcite{joustra2021} who found a significant advantage of later performances only in women’s gymnastics, which is plausibly driven by the existence of comparably more subjective evaluations in women’s competitions considering artistry.

\subsection{Fixed effects estimation}
While Figure~\ref{fig2} provides an insightful illustration of the effect of trimming on the reduction of nationalistic bias, it is important to consider idiosyncratic tendencies across judges (e.g., leniency or strictness), which may even vary \emph{within} a judge over years \parencite[e.g., due to experience; see][]{faltings2023, krumer2022}.

In the following, we use the continuous evaluation score assigned by each judge for a given performance as the unit of observation. Our data allow us to compare the scores of a compatriot judge with those of non-compatriot judges within the same run and then compare the deviations before and after the introduction of the trimming mechanism. More specifically, we estimate the following model: 
\begin{equation}\label{eq1}
Judge\ score_{jip} 
= \alpha_1 \, Compatriot_{jip} 
+ \alpha_2 \, (Compatriot_{jip} \times after_{jip}) 
+ \theta_p + \lambda_{js} + \varepsilon_{jip},
\end{equation}

where ${Judge\ score}_{jip}$ is the judge $j$’s score for run $p$ (performance) of athlete $i$. ${Compatriot}_{jip}$ is a dummy variable that receives the value of one if judge $j$ has the same nationality as athlete $i$, and zero otherwise. This variable identifies the average size of nationalistic bias before the change. ${Compatriot_{jip} \times after_{jip}}$ is an interaction term between the ${Compatriot}_{jip}$ variable and a dummy that receives the value of one for the period after the change (with trimming). $\theta_p$ is a run fixed effect. This means that we perfectly control for all characteristics that are constant within each run including athlete’s ability, her or his interim ranking in the competition, whether or not she or he competes in the home country, etc. Finally, to control for idiosyncratic tendencies across and within judges over years, we use judge-per-season fixed effects denoted by $\lambda_{js}$.

A positive (negative) sign of $\alpha_{1}$ implies a bias in favor of (against) a compatriot athlete before the introduction of the trimming mechanism. A negative (positive) sign of $\alpha_{2}$ implies a reduction (increase) in bias when the extreme scores are trimmed. A positive (negative) sign of the sum of $\alpha_{1}$ and $\alpha_{2}$ implies a bias in favor of (against) a compatriot athlete after the introduction of the trimming mechanism.

Table~\ref{tab3} presents the results of linear regressions based on equation~(\ref{eq1}), where each line represents the results for a different sample. In Line 1, we show the results for all the data combining both genders. We see that before the introduction of trimming ($\alpha_{1}$), judges assigned on average 0.365 points more to their compatriots than the other judges on panel that evaluated the same performance. This estimate is significant at the 1\% level. However, after the introduction of trimming, we see a relatively large reduction in nationalistic bias ($\alpha_{2}$) in size (about two thirds of the size of $\alpha_{1}$), though the effect is not significant at conventional levels. While the total size of nationalistic bias after the change ($\alpha_{1}$ + $\alpha_{2}$) is still positive, it is far from being statistically different from zero and, as said above, much lower in size than $\alpha_{1}$.

\begin{table}[!htbp]
\small
\centering
\caption{Fixed effects estimates of nationalistic bias before and after the introduction of the trimming rule across subsamples}
\label{tab3}
\resizebox{\linewidth}{!}{
\begin{tabular}{lrllrr}
\hline\hline
\makecell[l]{Sample}
 & \makecell{No. \\ of scores \\ (runs)}
 & \makecell{$\alpha_1$ \\ Nat. bias \\ without trimming}
 & \makecell{$\alpha_2$ \\ Compatriot\texttimes time \\ period}
 & \makecell{$\alpha_2$ \\ Change \\ in \%}
 & \makecell{$\alpha_1 + \alpha_2$ \\ Nat. bias \\ with trimming}
 \\
\hline
(1) All data (both genders)	& 29,383 (5,305) & \hspace{2mm} 0.365 (0.119)***	& \hspace{2mm} -0.240 (0.168) & -66\% & 0.125 [0.319] \\
\hline
(2) Men’s competitions (all) & 17,180 (3,100) &  \hspace{2mm} 0.343 (0.131)** & \hspace{2mm} -0.236 (0.174) & -69\% & 0.107 [0.349] \\
\hspace{5mm}(3) Scores$>$56 (median)	& 8,523 (1,538)	&  \hspace{2mm} 0.382 (0.126)***	& \hspace{2mm} -0.147 (0.214) & -38\% & 0.235 [0.148] \\
\hspace{5mm}(4) Scores$\leq$56 (median)	& 8,657 (1,562)	&  \hspace{2mm} 0.338 (0.167)** & \hspace{2mm} -0.344 (0.184)*	& -102\% & -0.006 [0.952] \\
\hspace{5mm}(5) Finals & 5,538 (990) &  \hspace{2mm} 0.635 (0.164)***	& \hspace{2mm} -0.530 (0.220)** & -83\% & 0.105 [0.493] \\
\hspace{5mm}(6) Qualifications & 11,642 (2,110) & \hspace{2mm} 0.214 (0.145) & \hspace{2mm} -0.078 (0.178) & -36\% & 0.136 [0.201] \\
\hline
(7) Women’s competitions (all) & 12,203 (2,205) & \hspace{2mm} 0.384 (0.162)** & \hspace{2mm} -0.213 (0.259) & -55\% & 0.171 [0.400] \\
\hspace{5mm}(8) Scores$>$64 (median) & 6,049 (1,088) & \hspace{2mm} 0.530 (0.183)*** & \hspace{2mm} -0.438 (0.225)* & -83\% & 0.092 [0.591] \\
\hspace{5mm}(9) Scores$\leq$64 (median) & 6,154 (1,117) & \hspace{2mm} 0.308 (0.190) & \hspace{2mm} -0.015 (0.380) & -5\% & 0.293 [0.404] \\
\hspace{5mm}(10) Finals	 & 4,784 (862) & \hspace{2mm} 0.477 (0.239)* & \hspace{2mm} -0.009 (0.495) & -2\% & 0.468 [0.217] \\
\hspace{5mm}(11) Qualifications & 7,419 (1,343) & \hspace{2mm} 0.387 (0.168)** & \hspace{2mm} -0.403 (0.247) & -104\% & -0.016 [0.928] \\
\hline\hline
\end{tabular}
}
\parbox{\linewidth}{\vspace{1mm}\footnotesize\noindent\justifying\textit{Notes:} Each line presents the results of a linear regression based on equation~(\ref{eq1}) for a given sub-sample. The dependent variable is the judge’s score for a given run. Standard errors are multi-way clustered at the athlete and judge level and appear in parentheses. Last column presents the sum of $\alpha_1 + \alpha_2$ that represents the nationalistic bias in the period with trimming. The \emph{p}-value of $H_0$: $\alpha_1 + \alpha_2 = 0$ is presented in square brackets. Significance levels are denoted as follows: * $\leq 10\%$, ** $\leq 5\%$, *** $\leq 1\%$.}
\end{table}

To provide a more nuanced picture, we stratify our sample by gender, performance scores, and competition phase. More precisely, within men’s and women’s competitions, we further divide the data into scores above and below the median to differentiate between performances that were most likely affected by falls (below the median) and those that were not and are therefore more relevant for the athletes‘ rankings (above the median). Finally, we distinguish between competition phases, because stakes are higher in the finals than in the qualifications.

In Lines 2-6, we show results for the men’s competitions. We see that in all sub-samples, the nationalistic bias without trimming ($\alpha_1$) is positive, and in four out of five cases it is also statistically significant at conventional levels. To put our results into perspective, before the introduction of the trimming mechanism, nationalistic bias accounts for 16\% of the within-run standard deviation in men’s competitions. In finals (the largest coefficient in the men’s sub sample), however, it accounts for as much as 29\% of the within-run standard deviation.\footnote{Note that the within-performance standard deviation for all men’s (women’s) data before the introduction of the trimming mechanism is 2.18 (2.28).} This is a sizable effect that is comparable to nationalistic bias found in other settings with standardized estimates ranging from 24\% in dressage to 47\% in figure skating \parencite[see Figure 5 in][]{krumer2022} . We also see a reduction in nationalistic bias in all sub-samples ($\alpha_2$), which is relatively large, and in two cases also statistically significant at conventional levels. Finally, in none of the sub-samples, nationalistic bias ($\alpha_1$ + $\alpha_2$) is statistically significant at conventional levels after the introduction of the trimming mechanism.

In Lines 7-11, we show results for the women’s competitions. Overall, we see a fairly similar picture as with the men’s data. In all sub-samples, the nationalistic bias without trimming ($\alpha_1$) is positive and in four out of five cases, it is also statistically significant at conventional levels. On average, the nationalistic bias accounts for 17\% of the within-run standard deviation in women’s competitions before the introduction of the trimming mechanism. We also see a reduction in nationalistic bias in all sub-samples ($\alpha_2$), which is relatively large except when the scores are below the median and in the finals. As in men’s competitions, in none of the sub-samples, nationalistic bias ($\alpha_1$ + $\alpha_2$) is statistically significant at conventional levels after the introduction of the trimming mechanism. However, while the bias is not statistically significant, its size is still relatively large particularly in the finals. Such a large size could be a random result, but it might also reflect some strategic considerations among the compatriot judge and the other judges in the panel. Especially in the finals, the stakes are high and evaluation could be more subjective than in men’s competitions.\footnote{Note that our results are robust to the exclusion of tournaments with seven judges. These results are available in the supplemental material.}

One might argue that the bias could technically be reduced even without the trimming mechanism and without any change in behavior of compatriot judges – simply because the size of the panel increased. To illustrate, for a panel with five judges that consists of a compatriot judge and the other four judges (judges 2-5), the bias is calculated as:

\begin{equation}\label{eq2}
Bias_{(if\ 5\ judges\ on\ panel)}
= score_{compatriot}
- \frac{\sum_{j=2}^{5} score_j}{4},
\end{equation}

while for a panel that consists of a compatriot judge and the other five judges (judges 2-6), the bias is calculated as:

\begin{equation}\label{eq3}
Bias_{(if\ 6\ judges\ on\ panel)}
= score_{compatriot}
- \frac{\sum_{j=2}^{6} score_j}{5}.
\end{equation}

No change in behavior of the existing judges (a compatriot and judges 2-5) means that they do not change their scores. In such a case, the greatest reduction in nationalistic bias that can be achieved, provided the compatriot judge assigns the highest score, occurs when the new judge (judge 6) assigns an identical score to the compatriot judge. Thus, we can re-write equation~(\ref{eq3}) as:

\vspace{-30pt}
\begin{equation}\label{eq4}
\begin{aligned}
Bias_{(if\ 6\ judges\ on\ panel)}
& = score_{compatriot}
- \left(\frac{\sum_{j=2}^{5} score_j + score_{compatriot}}{5}\right) \\
& = 0.8 \left(score_{compatriot}
- \frac{\sum_{j=2}^{5} score_j}{4}\right),
\end{aligned}
\end{equation}

suggesting that the maximal technical reduction in nationalistic bias by moving from a panel of five to a panel of six judges is 20\%. If, however, the score of judge 6 is lower than the score of the compatriot judge (which is plausible under the assumption of nationalistic bias), then the technical reduction in bias is less than 20\%.

According to our findings in Table~\ref{tab3}, the nationalistic bias is 66\% lower when looking at all the data (Line 1 in Table~\ref{tab3}), 69\% lower in the men’s competitions (Line 2 in Table~\ref{tab3}), and 55\% lower in the women’s competitions (Line 7 in Table 3). In fact, in nine out of eleven sub-samples presented in Table~\ref{tab3}, the reduction in nationalistic bias after the introduction of the trimming mechanism is much larger than 20\%. As such, and taking together all results – i.e., the reduction in the mean score of the compatriot judges (see Table~\ref{tab2}), the distribution of the score ranks, (see Figure~\ref{fig2}), and the results of our fixed effects estimations (see Table~\ref{tab3}) – we conclude that the reduction in nationalistic bias is mainly driven by the trimming mechanism rather than by the technical increase in the number of judges on panels.
\vspace{-10pt}
\section{Conclusion} \label{sec:conclusion}
Whether or not trimming the extreme votes can serve as a remedy for bias in subjective evaluations is, by and large, an empirical question. This is because different preferences (e.g., private in favor of a specific candidate and reputational in favor of sincere voting) may act in opposite directions. Motivated by the absence of evidence from real-world settings, this paper provides the first evidence outside of laboratory experiments about the effect of trimming on the behavior of professional experts whose subjects of evaluations are highly incentivized and professional individuals. 

Overall, our results suggest that trimming of extreme votes may not only mitigate biases ex post, by excluding the extreme votes from the overall panel count, but also ex ante, by changing the behavior of individual judges.

\printbibliography{\footnotesize}

\clearpage
\end{document}